\documentclass[aps,prd,amsmath
,eqsecnum            
,preprintnumbers
,nofootinbib         
,twocolumn         
]{revtex4-1}
\usepackage[dvips]{graphicx}
\usepackage{amsmath}
\usepackage{amsfonts}
\usepackage{amssymb}
\usepackage{longtable}   
\usepackage{color}
\usepackage{shapepar}
\usepackage{textcomp}
\allowdisplaybreaks[4]     

\begin{document}

\title{Thermodynamics and Lemaitre-Tolman-Bondi void models}

\author{Priti Mishra}
\email{priti@tifr.res.in}
\author{Tejinder P. Singh}
\email{tpsingh@tifr.res.in}
\affiliation{Department of Astronomy and Astrophysics,
Tata Institute of Fundamental Research, \\
Homi Bhabha Road, Colaba, Mumbai -  400005, Maharashtra, India}

\date { }

\begin{abstract}
\noindent It has been argued in the literature that in order to make a matter dominated Friedmann-Lemaitre-Robertson
-Walker universe compatible with the generalized second law of thermodynamics, one must invoke
dark energy, or modified gravity. In the present article we investigate if in a similar spirit, inhomogeneous 
cosmological models can be motivated on thermodynamic grounds. We examine a particular minimal void Lemaitre-Tolman
-Bondi inhomogeneous model which agrees well with observations. While on the one hand we find that the entropy associated 
with the apparent horizon is not well-behaved thermodynamically, on the other hand the canonical Weyl curvature 
entropy shows satisfactory thermodynamic behavior. We suggest that evolution of canonical Weyl 
curvature entropy might be a useful way to evaluate the thermodynamic viability of inhomogeneous cosmologies.
\end{abstract}
\maketitle 
\section{Introduction}
\noindent Supernovae Ia observations in 1998 first hinted at the accelerated expansion rate of the universe 
\cite{riess98, permutter99}. In a homogeneous and isotropic universe this accelerated expansion can be explained 
only by assuming a fluid with negative pressure. Such a fluid is referred to as dark energy. Ever since then there 
have been various candidates proposed for dark energy and also there have been various attempts to explain the 
accelerated expansion rate without invoking any dark energy component. The most popular and the simplest candidate 
for dark energy is cosmological constant ($\Lambda$). As the name suggests the energy density associated with 
cosmological constant does not evolve with time. One quantity which can play the role of cosmological constant 
is the vacuum because vacuum energy has the same equation of state as $\Lambda$. But the problem with $\Lambda$ is 
that the value required to match with observation is smaller than its predicted value of vacuum energy from quantum
 field theory by a factor of $10^{120}$. Therefore, the origin of cosmological constant is still unknown. And also 
the observed value of energy density of cosmological constant is of the same order as that of matter. This is called
 coincidence problem. There are other models of dark energy too namely, quintessence and k-essence models. In these 
models the dark energy evolves with space and time. Quintessence models solve the coincidence problem \cite{quintessence99,
 quintessence12,shruti12}. 
 k-essence models have non-standard form of kinetic energy. In these models the universe is likely to go through
 big rip which means that the scale factor would become infinite some time in future and the large scale structures would
 be ripped apart and in some quintessence models there would come a time for freeze out which means the structures would
 collapse into each other. 
 
Working within the framework of Einstein gravity dark energy is required only in homogeneous and isotropic universe. Since
 the scale of homogeneity is still not well established, it is perhaps interesting to relax this assumption of homogeneity 
of space and see whether dark energy is still required to explain the cosmological observations. 

Many exact inhomogeneous solutions of Einstein equations have been found. Only Lema\^itre-Tolman-Bondi (LTB), Szekeres and
 Stephani models have been studied in the context of cosmology. LTB is spherically 
 symmetric dust solution of Einstein equations \cite{lemaitre33, tolman34, bondi47}.  Szekeres is dust solution of Einstein 
equations  \cite{szekeres75}. Stephani metric is solution of Einstein equation when the energy density of the source has a 
pressure gradient \cite{stephani87}. There have been many inhomogeneous models proposed which attempt to explain the cosmological
observations without assuming any dark energy component \cite{tomita99,celerier00,iguchi02,alnes06,chung06,enqvist07,bolejko08a,
garciabellido08,garciabellido08b,zibin08,yoo08,enqvist08,Alexander09,BW09,CBK10,BNV10,MZS11,NaSa11,stephanimodel1,BC10, BS11}.

While the existence of dark energy was first hinted by SN Ia observations, there is a recent study which shows that
 for the generalized second law of thermodynamics to hold in Friedmann-Lema\^itre-Robertson-Walker (FLRW) universe a 
dark energy component or something equivalent in modified gravity models must exist \cite{Pavon2005, Pavon2012}. 
The authors of \cite{Pavon2012,Pavon2005} defined the gravitational entropy as sum of matter entropy and the entropy of the apparent horizon.
Apart from FLRW models, these authors  investigated various other cosmological models and found that Chaplygin gas models 
\cite{chaplygin1,chaplygin2}, some holographic dark energy models \cite{holographic1,holographic2}, Dvali, Gabadadze and 
Porrati (DGP) brane model \cite{DGP1,DGP2,DGP3} and Cardessian model \cite{cardessian} are allowed by generalized second
 law of thermodynamics. Therefore the existence of dark energy is demanded not only by observations but also by thermodynamics. 

Now it is interesting to see whether the inhomogeneous models which claim to explain the cosmological observations without dark
 energy satisfy the generalized second law of thermodynamics. For this we first need to define the gravitational entropy in 
inhomogeneous models. As a first step we follow the work of Pavon and Radicella by which we mean that we
 also define gravitational entropy as sum of the matter entropy and the entropy of the apparent horizon. But this definition may
 not be not universal because the apparent horizon may not exist in some inhomogeneous models while every gravitating system 
possibly has a gravitational entropy. Therefore, we explore other candidates of gravitational entropy as well. We consider 
Penrose's Weyl curvature hypothesis which proposes the relation between  the gravitational entropy and Weyl tensor \cite{penrose79}. 
Following this hypothesis many different forms of entropy have been proposed \cite{goodewainwright84,WaAnd84,GrHe01,HoBM04,Sussman13,
CET13}. And there have been various studies on the thermodynamics of various spacetimes \cite{tp_sukratu97,tp_sukratu97_essay,
TP_Pollock89,Cai2009, BolejkoStoeger13,Nairwita2011}. We choose few candidates of entropy and also a realistic inhomogeneous model. 
By realistic inhomogeneous model we mean that such an inhomogeneous model which provide a good fit to the data. Then we investigate 
whether these candidates can represent gravitational entropy for the realistic inhomogeneous models.

The plan of this paper is as follows: Section \ref{sec1} briefly explains Pavon and Radicella`s work related to thermodynamics 
of cosmological models. In section \ref{sec2} we briefly recall LTB models. Section \ref{sec3} describes the LTB void model for
 which we want to calculate the entropy. In section \ref{sec4} we explain the apparent horizon and calculate the entropy 
 of the LTB model. The results are shown in the same section.
In section \ref{sec5} various other candidates of gravitational entropy are calculated in LTB model and the results are shown and 
interpreted. Then we discuss the results in section \ref{sec6}.

\section{Thermodynamics and cosmological models}\label{sec1}

 The generalized second law of thermodynamics [GSLT] states that for a 
system to tend towards thermodynamic equilibrium its entropy should be ever increasing and its evolution should be of 
convex nature i.e. the entropy should satisfy the following two conditions: $S'>0$ and $S''<0$ \cite{callen}, where prime
 denotes an evolution parameter such as the scale factor or cosmological time. Pavon and Radicella [PR] have studied the 
thermodynamics of FLRW and some modified gravity models and argued that dark energy (or alternatively, 
modified gravity) is indicated on thermodynamic grounds \cite{Pavon2012}. Below we briefly describe PR's work: 

 PR proceed to calculate $S'$ and $S''$ in FLRW models as follows:
 The entropy of a FLRW universe is sum of the entropy of the apparent horizon and the entropy 
of the fluids enclosed by the horizon. They chose the entropy of apparent horizon because the apparent horizon in FLRW universes
always exists (which is not generally true for the particle horizon and the future event horizon) and is assumed to possess an 
entropy proportional to its area \cite{bak-rey,cai-2008} and also a temperature \cite{cai-2009}. Besides, it appears to be the 
appropriate thermodynamic boundary \cite{wang-2006}. Further, they assume that the entropy of the apparent horizon is proportional
to its area via the following equation:
\begin{equation}
 S_{A} \equiv \frac{k_{B}}{4}\,\frac{\cal A }{l _{Pl}^2} 
\end{equation}
where $l_{Pl}$ and $k_{B}$ are Planck length and Boltzmann constant respectively.
 ${\cal A}$ is the surface area of the apparent horizon and can be calculated from 
 \begin{equation}
  {\cal A} = 4 \pi \tilde{r}_{A}^{2}
 \end{equation}
where 
\begin{equation}
\tilde{r}_{A}= (\sqrt{H^{2} + (k/a^{2})})^{-1}.
\end{equation}
 $\tilde{r}_{A}$ is the radius of the apparent horizon and $H$ the Hubble factor of the FLRW metric \cite{bak-rey}. 

Using Friedmann equations and the conservation equation it can be shown that 
\begin{equation}
{\cal A} = 4 \pi \, \tilde{r}_{A}^{2} = \frac{3}{2 G}\,
\frac{1}{\rho} \, ,\label{harea}
\end{equation}
and
\begin{equation}
{\cal A}' = \frac{9}{2G} \, \frac{1\, +  \,w}{a \, \rho} \, ,
\label{primeA1}
\end{equation}
where $w$ is the equation of state parameter $p=w\rho$. Eqn. (\ref{primeA1}) shows that the area will increase 
in expanding universes if $1+ w > 0$.

For $w =$ constant Eqn. (\ref{primeA1}) reduces to 
\begin{equation}
{\cal A}'' = - \frac{9}{2G}\, \frac{1 \, + \, w}{(a \,
\rho)^{2}}\, (a\, \rho' \, + \, \rho) =  \frac{9}{2G \, a^{2}\,
\rho } \, (1 \, + \, w) \, (2 \, +\, 3w) \, . \label{2primeA1}
\end{equation}
Above equation shows that $ {\cal A}'' \leq 0$ for $-1 \leq w \leq -2/3$. Therefore, the FLRW models with equation of 
parameter in this range are not favored by the convexity condition of GSLT.

The entropy of the fluid enclosed by the apparent horizon is calculated from following equation:
\begin{equation}
 S_{m} = k_{B} N 
\end{equation}
where $N$ is the number of dust particles contained within the apparent horizon. 
 This equation calls for an explanation. As noted by \cite{Pavon2012}, since one is dealing with 
pressureless matter (i.e. dust), the matter is cold, and has an effective temperature $T=0$, which does not permit the 
definition of an entropy in a traditional manner. Hence they propose associating one unit of entropy ($k_B$) with each
particle. We can also motivate this somewhat differently, by considering the fundamental relation
\begin{eqnarray} 
TdS &&= dU + P dV \\
&&= k_B T dN + PdV\\
&&= k_B T dN + (\rm neglibile\ \rm term)
\end{eqnarray}
at a constant temperature $T$, and by noting that because the matter is non-relativistic [dust], the pressure is extremely 
small, and the pressure term may be ignored. Then by comparing the left and right hand side, and treating the temperature 
as a very small but non-zero quantity which cancels out,   Eqn. (2.7) follows, being the same relation as proposed by
\cite{Pavon2012}. 
In homogeneous and isotropic universe $N$ 
should have the following form:
\begin{eqnarray}
 \, N = (4\pi/3)\tilde{r}_{A}^{3} n 
\end{eqnarray}
where $n= n_{0} \, a^{-3}$ is the number density of dust particles. Substituting the form of $N$ into the expression of 
matter entropy we obtain 
\begin{equation} 
S_{m} =  k_{B}\,  \frac{4 \pi}{3}\, \tilde{r}_{A}^{3} \, n_{0}\,
a^{-3}  \label{eq:Smatter1}
\end{equation} 
which shows that the matter entropy is proportional to $ a^{3/2}$.  
Hence $S''_{m}$ is positive. It shows that $S''_{m} \, + \, S''_{A} > 0$ in matter dominated FLRW universe,
 which is not in accordance with the GSLT.
 
 Thus it has been shown by PR that in a pure matter FLRW universe the total entropy which is 
the sum  of matter entropy and the entropy of the apparent horizon does not satisfy the convexity condition. 
When a dark energy component is added to the FLRW universe the entropy satisfies the generalized
 second law of thermodynamics. This dark energy can be the cosmological constant or a evolving dark energy i.e. 
quintessence or k-essence models. 
 
 PR also studied the thermodynamics of modified gravity models. They found that Chaplygin gas model 
\cite{chaplygin1,chaplygin2}, some holographic dark energy models \cite{holographic1,holographic2}, Dvali, Gabadadze 
and Porrati (DGP) brane model \cite{DGP1,DGP2,DGP3} and Cardessian model \cite{cardessian} also is in keeping with the 
generalized second law of thermodynamics.

\section{LTB models}\label{sec2}
The LTB metric is the spherically symmetric solution of Einstein equations for dust source. In comoving coordinates the LTB 
metric is given by:
\begin{equation}
{\rm d} s^2 =  c^2 {\rm d} t^2 - \frac{\Phi'^2}{1 - k} {\rm d} r^2 - \Phi^2 ({\rm d} \theta^2 + \sin^2\theta{\rm d} \phi^2),
\label{ltbmetric}
 \end{equation} 
 where $\Phi(t,r)$ is the area distance to the comoving shell at distance $r$ from the origin and $k(r)$ is the curvature function.

Applying the LTB metric to the Einstein equations we obtain following equation:
 \begin{equation} 
\frac{1}{c^2}\dot{\Phi}^2 = \frac{2M}{\Phi} - k. \label{einsteineqninltb}
\end{equation} 
and 
\begin{equation} 
\kappa \rho c^2= \frac{2M' }{\Phi^2 \Phi' }, \label{rho}
\end{equation} 
where $\kappa=8\pi G/c^4$ and $M(r)$ is the gravitational mass within the comoving shell of radius $r$. 
 
The Einstein Eqn. (\ref{einsteineqninltb}) has following solutions depending upon the sign of the curvature function $k(r)$ :
\indent \newline \newline 
Case 1. $k<0$ 
\begin{eqnarray} 
\Phi=\frac{M}{(-k)}(\cosh\eta-1),\\
c(t-t_B(r))=\frac{M}{(-k)^{3/2}}(\sinh\eta-\eta)\label{knegative}
\end{eqnarray} 
\indent \newline \newline 
Case 2. $k>0$
\begin{eqnarray} 
\Phi=\frac{M}{k}(1-\cos\eta),\\
c(t-t_B(r))=\frac{M}{k^{3/2}}(\eta-\sin\eta)\label{kpositive}
\end{eqnarray}
\indent \newline \newline 
Case 3. $k=0$
\begin{equation} 
\Phi=\left(\frac{9M}{2}\right)^{1/3}\left(t-t_B(r)\right)^{2/3}\label{kzero}
\end{equation} 
where $t_B(r)$ is called the bang time function. It tells the time of big bang for worldlines of comoving radius $r$. The case 
when $k$ is positive, the evolution is called elliptic evolution because in this case the universe is closed which means that 
the the expansion of the universe would stop and after sometime it will start collapsing. The case when $k$ is negative, the 
evolution is called parabolic. In this case universe is open and it will be ever expanding or ever collapsing. The last case
when $k$ is zero, the universe is flat which means that the expansion or the contraction will be everlasting. The LTB metric
will yield FLRW metric if the mass density is homogeneous. 

In LTB model there are three arbitrary functions out of which one is fixed by the coordinate choice since the equations so far are covariant
under coordinate transformations.
\section{Minimal LTB void model}\label{sec3}
We are interested in studying the evolution of the entropy in a realistic LTB void model which is a good fit to observations. 
There is one such minimal void model by Alexander et al. (AMV) which is good fit to SN Ia data and is also consistent with the
WMAP 3-year data and the local measurements of Hubble parameter \cite{Alexander09}. In this LTB void model the mass function, 
the curvature function and bang time function are defined as follows: 
given by 
\begin{equation} 
M(r)=\frac{1}{6}\bar{M}^2M_P^2 r^3,
\end{equation} 
\begin{equation} 
k(r)=-2(\bar{M}r)^2k_{max}\left[1-\left(\frac{r}{L}\right)^4\right]^2,
\end{equation}
\begin{equation} 
t_B(r)=0.
\end{equation}
where $ M_P$ is the Planck mass, $\bar{M}$, $k_{max}$ and $L$ are parameters in the model. $L$ is also the size of the LTB void 
beyond which the universe is described by FLRW metric. For the best fit, $L$ is 250 Mpc$h^{-1}$ where $h$ is .55 .
$\bar{M}$ is determined from
\begin{equation} 
\bar{M} = \frac{h_{out}}{3000} \sqrt{\frac{3}{8\pi}}
\end{equation} 
where $h_{out}$ is the Hubble parameter in the FLRW region. 
$k_{max}$ is a parameter related to the density contrast at the center of the LTB void. 
\section{Thermodynamics of AMV model}\label{sec4}
Motivated by the work of PR we wish to see whether AMV model  
is in agreement with GSLT. For this we will calculate the gravitational entropy in this model and 
see whether the entropy satisfies both the conditions of GSLT. As a first step, following PR we also define 
entropy as sum of the entropy of the apparent horizon and the matter entropy. Before calculating the entropy of 
apparent horizon, we briefly recall the definition of apparent horizon and its mathematical expression for LTB 
models.
\subsection{Apparent horizon}
Since we want to calculate the entropy of apparent horizon, it is necessary to first define what an 
apparent horizon is and what is the mathematical condition for it to exist in a LTB universe. An apparent horizon is 
the outer envelope 
of a region of closed trapped surfaces. A closed trapped surface S$_{t}$ is such a surface from
which it is impossible  to send a diverging bundle of light rays. At this surface both the outward-directed
 and the inward-directed bundles immediately converge. Mathematically, this statement can be written as:  
 On S$_{t}$
 \begin{equation}
 k^\mu_{;\mu}\leq 0 
 \end{equation}
 where $k^\mu$ is the tangent vector to the surface. 
 
 The mathematical condition for the apparent horizon in LTB spacetime is \cite{kras_pleb2006}:
\begin{equation} 
\Phi=2M.\label{apparenthorizoncondition}
\end{equation} 

 After having written the mathematical expression of the apparent horizon in LTB models, we proceed 
 to find to calculate the apparent horizon in AMV model using the following algorithm: 
\subsection{Algorithm to find the Apparent Horizon in AMV Model} 
Since AMV model contains two different spacetimes in the range $r<L$ and $r>L$, we will calculate the apparent horizon 
in three steps: i) $r<L$, ii) $r=L$ and iii) $r>L$ where $L$ is the size of the void. 
\indent \newline \newline 
\textbf{Case 1: $r<L$ (Inside the void):}
\begin{enumerate}
 \item Since the curvature function $k(r)$ in AMV model is negative inside the void, the parametric solution of the Einstein 
 equations in open LTB spacetime is given by 
\begin{eqnarray} 
\Phi=\frac{M}{(-k)}(\cosh\eta-1)\label{parasola},\\
c(t-t_B)=\frac{M}{(-k)^{3/2}}(\sinh\eta-\eta)\label{parasolb}
\end{eqnarray}
where $\eta$ is a parameter.
\item Since the area radius of the apparent horizon in LTB spacetime follows the condition given in eqn. (\ref{apparenthorizoncondition}), 
the above two equations reduce to the following two equations at $r=r_A$:
\begin{eqnarray} 
-2k=\cosh\eta-1\label{parasola1},\\
c(t-t_B)=\frac{M}{(-k)^{3/2}}(\sinh\eta-\eta).\label{parasolb1}
\end{eqnarray} 
\item We then eliminate $\eta$ from eqns. (\ref{parasola1}) and (\ref{parasolb1}) to find the following equation:
\begin{eqnarray} 
c(t-t_B)&=&\frac{M}{(-k)^{3/2}}\left(\sinh\left(\cosh^{-1}\left(1-2k\right)\right)\right. \nonumber \\
&-&\left. \cosh^{-1}\left(1-2k\right)\right)
\end{eqnarray} 
Above equation can also be written as:
\begin{eqnarray} 
c(t-t_B)&=&\frac{M}{(-k)^{3/2}}\bigg[\left(\sqrt{\left(1-2k\right)^2-1}\right) \nonumber \\ 
 &-&  \cosh^{-1}\left(1-2k\right)\bigg]\label{raeqn1}
\end{eqnarray} 
\item  In the above equation we substitute the functions $M(r)$, $k(r)$ and $t_B$ for AMV model and obtain 
the time at which the apparent horizon lies at $r=r_A$ Mpc. 
\end{enumerate}
However, the above algorithm will work only when $r<L$ where $L$ is the size of the LTB void because the sign of 
the curvature function $k(r)$ inside the void is different from that at the boundary of the void. Hence, the solution
to the Einstein equations will also be different at the boundary. 

Therefore, we use the following algorithm to calculate the apparent horizon at the boundary of the LTB void:
\indent \newline \newline 
\textbf{Case 2: $r=L$ (At the boundary of the void):}
\begin{enumerate}
 \item Since the curvature function is zero at the boundary, the area radius at the boundary is obtained from the 
 following equation which is the solution to the Einstein equation for LTB spacetime with curvature function $k(r)$ 
 as zero:
\begin{equation} 
\Phi=\left(\frac{9M}{2}\right)^{1/3}\left(ct\right)^{2/3}.\label{phiatboundary}
\end{equation} 
\item Alternatively, one can also find the area distance by solving the Einstein equations for flat FLRW spacetime because
 $r=L$ is the interface between LTB and FLRW spacetimes in the AMV model. The solution to flat matter dominated FLRW tells
 that the area distance at the boundary is:
\begin{equation} 
a(t) r= \left(\frac{3}{2}H_0\sqrt{\Omega_m}t\right)^{2/3}r.
\end{equation} 
\item We substitute $\Phi$ from Eqn. (\ref{phiatboundary}) into the Eqn. (\ref{apparenthorizoncondition}) and obtain the 
following equation:
\begin{equation} 
t=\frac{4}{3c}M(r).
\end{equation} 
We calculate the apparent horizon at the boundary using above equation.
\end{enumerate}
\indent \newline \newline 
\textbf{Case 3: $r>L$ (Outside the void):}
\begin{enumerate}
 \item In AMV model the universe is described by flat FLRW metric outside the LTB void. Therefore, we will use the  solution of 
 the matter dominated FLRW equation to find the apparent horizon. In matter dominated flat FLRW universe the scale factor is given by
\begin{equation} 
a(t)=\left(\frac{3}{2}H_0\sqrt{\Omega_m}t\right)^{2/3}.
\end{equation}
 \item We substitute the functional form of the scale factor into the following equation: 
\begin{equation} 
a(t) r= 2 M(r)\label{app_hor_con}
\end{equation} 
which is same as Eqn. (\ref{apparenthorizoncondition}) with $\Phi$ written in the form of its FLRW limit.
\item After substituting the scale factor into Eqn.(\ref{app_hor_con}), we obtain the following equation:
\begin{equation} 
t=\left(\frac{2M(r)}{r}\right)^{3/2}\frac{2}{3 H_0 \sqrt{\Omega_m}}.
\end{equation}  
The above equation will give the radius of apparent horizon as a function of time in the FLRW region.
\end{enumerate}
\subsection{Entropy in Minimal LTB Void Model}
 After having found the area radius of apparent horizon, we calculate the gravitational entropy using the following algorithm:
 \begin{enumerate}
  \item  Following PR we also calculate the entropy of the apparent horizon from the following equation:
 \begin{equation} 
 S_A=\frac{k_B}{4} \frac{\cal{A}}{l_{Pl}^2}=\frac{k_B}{4} \frac{4\pi \Phi^2}{l_{Pl}^2},
 \end{equation} 
where $\cal{A}$ is the surface area of the apparent horizon and $l_{Pl}$ is Planck length. 
\item Then we calculate the matter entropy as follows: 

If there are $N$ dust particles each of mass $m$, the matter entropy of the
 system would be given by
\begin{equation}
 S_m=k_B N \label{mat_En}
\end{equation}
\item Assuming that the dust particles are non-interacting and dividing the total mass contained within the apparent
horizon by the mass of the dust particle we obtain the number of particles as 
\begin{equation}
 N=\frac{Mc^2/G}{m}
\end{equation}
where $m$ is the mass of the dust particle and $M(r)c^2/G$ is the total gravitational mass contained within the comoving 
shell of radius $r$. 
\item Substituting $N$ from above equation into the Eqn. (\ref{mat_En}) we obtain 
\begin{equation} 
S_m=k_B \frac{Mc^2/G}{m}\label{mattentropy}
\end{equation} 
\item We assume $m=10^{10}$ M$_\odot$ because it is mass of a typical galaxy and galaxy is considered to play the role of 
dust particle at cosmological scales. However, we find that our results do not change when we vary $m$ in a significant 
range around this value.
 \end{enumerate}

 \subsection{Results}
 \begin{figure}[!htb]
\includegraphics[width=9.5cm]{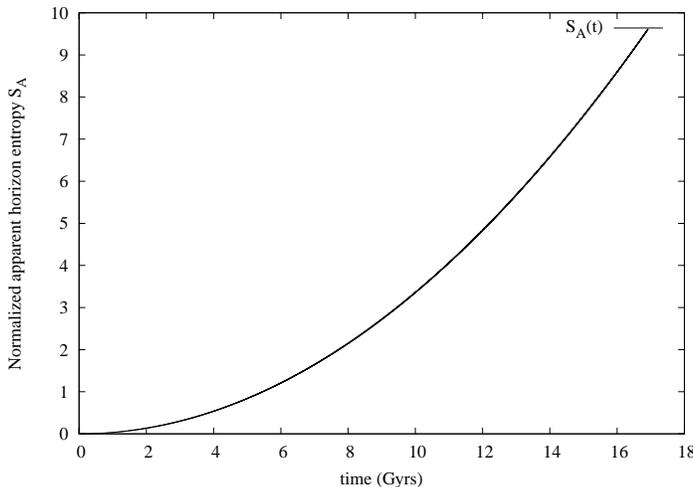}
\caption{Entropy of the apparent horizon vs. time where normalization factor is $10^{99}$ J/K.}
\label{s_avst}
\end{figure}

\begin{figure}[!htb]
\includegraphics[width=9.5cm]{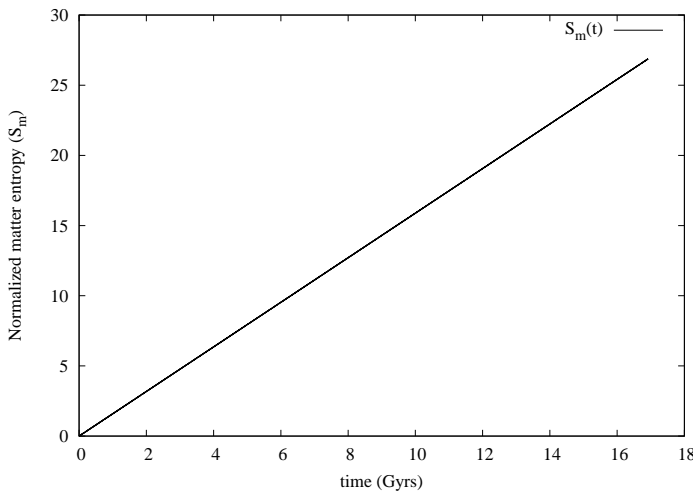}
\caption{Matter entropy vs. time where normalization factor is $10^{-12}$ J/K.}
\label{s_mvst}
\end{figure}
 Fig. \ref{s_avst} shows that the entropy of the apparent horizon is concave in nature. Fig. \ref{s_mvst} 
displays the behavior of the matter entropy with time. Fig. \ref{s_mvst} shows that the matter
 entropy increases almost linearly with time. This linear behavior was expected also. The matter entropy
 is proportional to the mass contained within the apparent horizon. According to Eqn. (\ref{apparenthorizoncondition})
 this mass is proportional to the area radius radius of the apparent horizon which increases linearly with
 time. Hence, the double derivative of the matter entropy will be very small compared to that of the entropy 
of the apparent horizon (Fig. \ref{s_mvst}). The total entropy is the sum of the matter entropy and  the 
entropy of the apparent horizon. Since the apparent horizon entropy is of concave nature and double derivative
 of the matter entropy is very small, the sum of these two entropies will also be of concave nature. 
 
 Therefore, according to this definition of entropy this minimal void model does not satisfy the second condition
 of the generalized second law of thermodynamics. One could hence take the stance that the model is anti-thermodynamic 
 or this definition is not a good representative of entropy in this minimal void model.
 
 Another approach however would be to argue that inhomogeneous cosmological models which do not possess spatial symmetry 
 may not have an apparent horizon at all. Keeping such a situation in mind, one could advocate looking for a more generic 
 measure of gravitational entropy, such as Weyl curvature. We investigate this next. 
 
\section{Weyl tensor and gravitational entropy }\label{sec5}
The Weyl curvature hypothesis was proposed by Penrose.  The starting point for the Weyl curvature hypothesis is the 
observation that the Big Bang singularity was very special. Penrose argues that the universe must have been in a low 
entropy state initially in order for there now to be a second law of thermodynamics. Assuming, as is commonly done, 
that the matter content of the universe was in thermal equilibrium near to the big bang, and therefore in a state of
 high entropy, one needs the contribution to the entropy from the rest of physics, which means the contribution from
 gravity or equivalently geometry, to be low. That is, the geometry must be highly ordered. Penrose introduced the 
concept of gravitational entropy and proposed that it is related to Weyl tensor. This hypothesis is called Weyl curvature
 hypothesis. The reason for proposing that the gravitational entropy is related to Weyl tensor because Weyl tensor 
is zero in a homogeneous state and it is non-zero for a universe with inhomogeneous structures.Therefore, if the 
gravitational entropy is related to Weyl tensor, it would evolve from a low entropy state to high entropy state which
 would be in accordance with the second law of thermodynamics. 
 
Following the proposal of this hypothesis many candidates of the gravitational entropy have been proposed. 

\subsection{Candidates for gravitational entropy}
We consider the following candidates for gravitational entropy motivated by the discussion in 
\cite{BolejkoStoeger13}:
\begin{enumerate}
\item 
\begin{equation} 
S= k_B \, l_{Pl} \int {\rm d}^3x \sqrt{h} C\label{entropy_def0}
\end{equation}
where $C$ is Weyl scalar which is defined as 
\begin{equation} 
C=C^{abcd}C_{abcd},
\end{equation} 
 $h$ is the determinant of the spatial part of the metric.  Here and in the following, the fundamental length unit Planck length 
 $l_{Pl}$ has been introduced so as to obtain correct dimensions for entropy.
\item The standard canonical definition \cite{WaAnd84}
\begin{equation} 
\delta s_c = k_B \frac{C_{abcd} C^{abcd}}{{\cal R}_{ab} {\cal R}^{ab}},
\label{s_c}
\end{equation} 
where $C_{abcd}$ is the Weyl tensor, and  ${\cal R}_{ab}$ is the Ricci tensor.
\begin{eqnarray}  
  R_{ab}R^{ab} =\frac{4 M'^2}{\Phi'^2\Phi^4}.\label{r_abr^ab}
\end{eqnarray} 
and 
\begin{equation} 
C=C_{abcd}C^{abcd}=\frac{12}{\Phi^4}\left(\frac{2 M'}{3\Phi'}-\frac{2 M}{\Phi}\right)^2\label{weylscalar}.
\end{equation} 
Substituting eqns. (\ref{r_abr^ab}) and (\ref{weylscalar}) into the Eqn. (\ref{s_c}) we obtain
\begin{equation}  
\delta s_c= k_B \frac{12\Phi'^2}{M'^2}\left(\frac{M'}{3\Phi'}-\frac{M}{\Phi}\right)^2.
\end{equation} 
\item The integrated version of the canonical definition:
\begin{eqnarray}  
S_c &=&  k_B/l_{Pl}^3 \int {\rm d}^3x \sqrt{h} \delta s_c=\int {\rm d}^3x \sqrt{h} \frac{C}{{\cal R}_{ab} {\cal R}^{ab}}\\
&=&4\pi k_B/l_{Pl}^3\int_{0}^{L} dr \left\{\frac{\Phi^2 \Phi'}{\sqrt{1-k}}\frac{12\Phi'^2}{M'^2}\left(\frac{M'}{3\Phi'}
-\frac{M}{\Phi}\right)^2\right\}.
\label{int_s_c}
\end{eqnarray}  

\item The third candidate is the canonical definition multiplied by the square root of the determinant of
 the spatial metric \cite{GrHe01},

\begin{equation} 
S_h =  k_B/l_{Pl}^2\sqrt{h} \frac{C_{abcd} C^{abcd}}{{\cal R}_{ab} {\cal R}^{ab}}=\frac{\Phi^2 \Phi'}{\sqrt{1-k}}
\frac{12\Phi'^2}{M'^2}\left(\frac{M'}{3\Phi'}-\frac{M}{\Phi}\right)^2 .
\label{S_GrHe}
 \end{equation} 
We consider this candidate because it is shown in \cite{GrHe01} that it is in accordance with the second law 
of thermodynamics. We investigate whether it can represent gravitational entropy in AMV model.
 \end{enumerate}
 
 \subsection{Results}
 \begin{figure}[!htb]
\includegraphics[width=9.5cm]{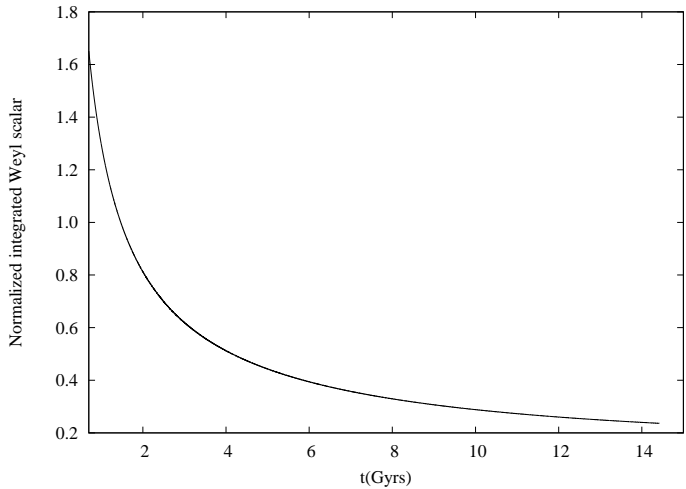}
\caption{Integrated Weyl scalar vs. time where normalization factor is  $10^{-86}$ J/K.}
\label{entropy_0}
\end{figure}

\begin{figure}[!htb]
\includegraphics[width=9.5cm]{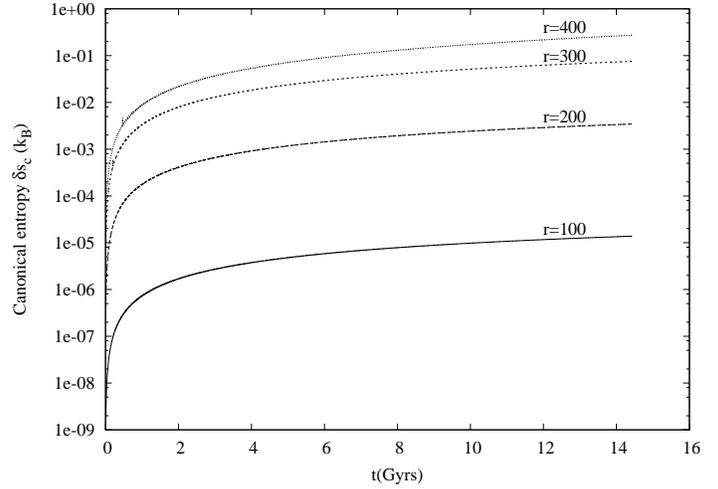}
\caption{Entropy ($\delta s_c$) vs. time where $r$ is in Mpc.}
\label{entropy_1}

\end{figure}
\begin{figure}[!htb]
\includegraphics[width=9.5cm]{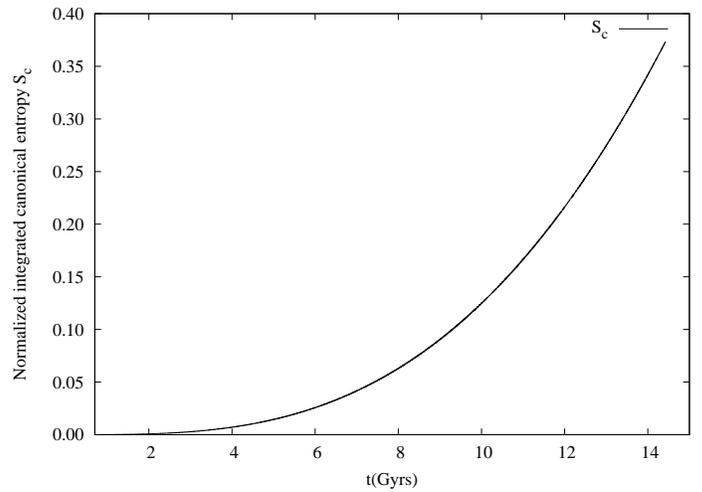}
\caption{Entropy $S_c$ vs. time where normalization factor is $10^{157}$ J/K.}
\label{entropy_2}
\end{figure}

\begin{figure}[!htb]
\includegraphics[width=9.5cm]{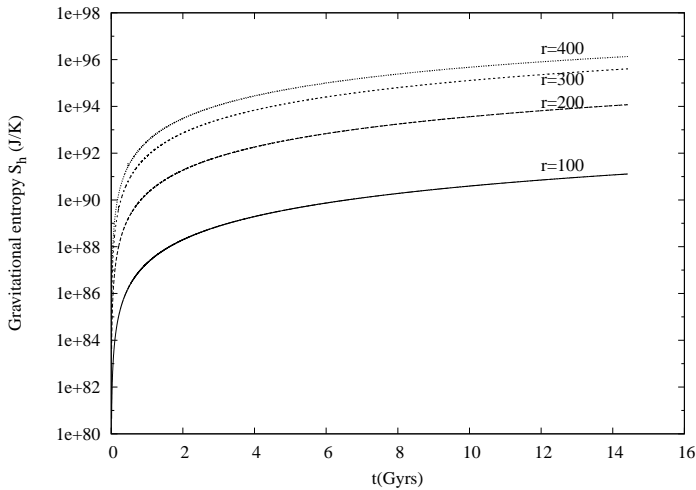}
\caption{Entropy $S_h$ vs. time where $r$ is in Mpc.}
\label{entropy_3}
\end{figure}

Fig. \ref{entropy_0}  depicts the behavior of the first candidate of entropy Eqn. (\ref{entropy_def0}) with time. 
It shows that the entropy according to this definition  is ever decreasing with time while 
we know that the entropy is ever increasing. Hence, the first candidate does not represent entropy for Alexander 
et al. LTB model. Figs. \ref{entropy_1}-\ref{entropy_3} display the time evolution of the other three candidates
 of entropy (eqns.(\ref{s_c}), (\ref{int_s_c}) and (\ref{S_GrHe})). In Figs. \ref{entropy_1} and \ref{entropy_3} 
entropy has been plotted for many different values of $r$ and all of them show same behavior.

We calculated four different candidates of the gravitational entropy proposed in the literature. We find that the 
first definition which is the integral of Weyl scalar over the volume of the void decreases with time. Hence, it 
does not represent the gravitational entropy. The second candidate which is the canonical definition of entropy 
increases linearly with time. When we add this entropy to matter entropy, the total entropy satisfies both condition
 of the second law of thermodynamics. Hence, the canonical definition represents the gravitational entropy in this
 LTB model. The third candidate is the integral of the first candidate over the volume of LTB void. This entropy 
increases with time but the entropy is of concave nature which does not satisfy the second condition of the 
GSLT which demands the convexity of entropy. Hence, the third 
candidate also does not represent the gravitational entropy. The fourth candidate also increases with time but 
it is not of convex nature. Hence, this candidate also does not represent the gravitational entropy. 

Therefore, the canonical definition of Weyl entropy is a good representative of the gravitational 
entropy in the AMV model. 

We would like to propose that further investigations of realistic inhomogeneous cosmological models
 should be carried out, to investigate the behavior of the canonical Weyl entropy. Such studies could 
help discriminate between different inhomogeneous models on thermodynamic grounds.

\section{Discussion}\label{sec6}
We considered following five candidates for gravitational entropy:
\begin{enumerate}
 \item The sum of the entropy of apparent horizon and matter entropy.
\item The integrated Weyl scalar over the LTB volume.
\item The canonical definition. 
\item Integrated canonical entropy over LTB volume.
\item The product of canonical entropy by the square root of determinant of spatial metric.
\end{enumerate} 
We calculated these candidates of entropies for AMV model and found the following results.
\begin{enumerate}
\item The entropy of the apparent horizon is positive and increases with time. But its time evolution is concave in nature. The
 matter entropy increases almost linearly with time. Therefore, the sum of these two entropies would be of concave nature. Here, 
one should note that the matter entropy depends upon the mass of the dust particle. However, since the mass of the dust particle 
is just a multiplicative factor in the expression of matter entropy, its concave or convex nature will not change regardless of 
the mass of dust particle because the matter entropy increases linearly with time. Thus, we can infer that this
candidate is not in keeping with second condition of GSLT. Therefore, it cannot be gravitational entropy for AMV model. However, we do 
not rule out this candidate for all inhomogeneous models. One should test this candidate for other models as well.
\item The integral of Weyl scalar over AMV volume is positive but decreases with time. It does not fulfill the first condition of 
GSLT which requires the entropy to be ever increasing. Hence, integrated Weyl scalar does not represent gravitational entropy in AMV 
model.
\item The canonical definition of gravitational entropy varies with comoving distance as well as with time. Therefore, we fix different 
values of $r$ and study its evolution. We find that it shows same behavior for four different values of $r$ in the large range 100-400
 Mpc at interval of 100 Mpc. Hence, we can infer that the behavior of evolution of canonical entropy does not change 
 with cosmological distance. Canonical entropy is positive and ever increasing. The rate of increase is almost linear and hence it fulfills
 the convexity condition of GSLT. 
 Therefore, when this is added to the canonical entropy, the total entropy increase  is in accordance 
 with GSLT. Therefore, 
 it can be a good representative of gravitational entropy in AMV model. However, this candidate should be tested for other inhomogeneous 
 models as well to put stringent constraints on its being gravitational entropy candidate.
\item The integrated version of canonical definition of entropy is positive and is ever increasing also. However, the rate of increase is of 
concave nature. Hence, this candidate is not in keeping with the second condition of GSLT. Even when added to matter entropy its evolution will 
not be of convex because the matter entropy increases linearly with time and hence its second derivative with respect to time is zero and 
therefore it will not alter the overall behavior of total entropy. The nature of only the gravitational entropy candidate will decide whether 
the total entropy is concave or convex. 
\item The last candidate which we studied is the product of canonical entropy and the square root of the determinant of spatial metric. This 
candidate also depends upon time as well as on the comoving distance $r$. We fix four different values of $r$ in the range 100-400 Mpc at interval
 of 100 Mpc and study its evolution. We find that for all these values this candidate shows same behavior. Hence, the evolution of 
this candidate does not depend on the cosmological distance. It is positive and increases monotonically. However, its evolution is of concave nature 
and hence it does not fulfill the convexity condition of GSLT. Therefore, this candidate also does not represent the gravitational entropy in AMV 
model. 
\end{enumerate}
We summarize our results by saying that out of all the five candidates of entropy only canonical definition fulfills all conditions of 
GSLT in AMV model. This candidate should further be studied for other inhomogeneous models. 

Assuming that canonical definition is the correct definition of gravitational entropy AMV model is in keeping with the 
generalized second law of thermodynamics. Hence, dark energy is not required in a LTB void model by generalized second law of thermodynamics 
while it (or some alternative) is required in FLRW models (or modified gravity models).

\bigskip 
{\it Note added in proof}. -- It has been brought to our attention
\cite{zibin} that unlike what we stated at the start of Sec. IV above,
the full analysis of the CMB spectrum in the AMV model
was carried out in \cite{BNV10}, and that the fitted local Hubble
parameter is extremely low. Observational constraints on
LTB models have been studied in \cite{zibin08}, \cite{MZS11} and also in
\cite{marra2010}, \cite{bull2012}. While we agree that there are very strong
observational constraints on LTB void models, one can
nonetheless explore their thermodynamic properties, as has
been done by us in this paper. It was pointed out to us \cite{larena} 
that gravitational entropy in LTB dust models has been
studied also in \cite{sussmanlarena2014} using the Clifton-Ellis-Tavakol and
Hosoya-Buchert proposals for entropy.

 \begin{acknowledgments}
 The tensor calculations were done using mathtensor package of Mathematica. We would like to thank James P.
Zibin, Julien Larena, Roberto Sussman, and Thomas
Buchert for their useful comments on our paper, and for
bringing Refs. \cite{marra2010}, \cite{bull2012}, \cite{sussmanlarena2014} and \cite{li2012} to our attention.

 \end{acknowledgments}

\end{document}